\documentclass[
aps,prb,floats,showpacs,eqsecnum,amsmath,graphicx,
amsmath,amssymb,twocolumn]{revtex4}
\usepackage[dvips]{graphicx}
\usepackage{epsfig,color}
\begin{document}
\title{As-vacancies, local moments, and Pauli limiting in
LaO$_{0.9}$F$_{0.1}$FeAs$_{1-\delta}$ superconductors}
\author{Vadim Grinenko$^{1}$, Konstantin Kikoin$^{1,2}$, 
Stefan-Ludwig Drechsler$^1$ \footnote{Corresponding author, E-mail: s.l.drechsler@ifw-dresden.de},
G\"unter Fuchs$^1$, Konstantin Nenkov$^{1,3}$, \\
Sabine Wurmehl$^1$,
Franziska Hammerath$^1$, Guillaume Lang$^1$, Hans-Joachim
Grafe$^1$,\\ Bernhard Holzapfel$^1$, Jeroen van den Brink$^1$, Bernd
B\"uchner$^1$, and Ludwig Schultz$^1$}
   \affiliation{
$^1$Leibniz Institute for Solid State and Materials Research
IFW--Dresden, P.O. Box 270116, D-01171 Dresden,
Germany\\
$^2$ School  of Physics and Astronomy, Tel-Aviv University,
Tel-Aviv 69978, Israel\\
$^3$ International Laboratory of High Magnetic Fields and Low
Temperatures, Gajowicka, 95, PL-53-529 Wroclaw, Poland
   }
\begin{abstract} {
We report magnetization measurements of
As-deficient ${\rm LaO_{0.9}F_{0.1}FeAs_{1-\delta}}$
($\delta \approx 0.06$) samples with improved superconducting
properties as compared with As-stoichiometric optimally doped
La-1111 samples. In this As-deficient system
with almost homogeneously distributed As-vacancies (AV),
as suggested by the $^{75}$As-nuclear quadrupole resonance (NQR) measurements,
we observe a strong
enhancement of the spin-susceptibility by a factor of 3-7. This
observation is attributed to the presence of an electronically
localized state around each AV, carrying a magnetic moment of
about 3.2 $\mu_B$ per AV or 0.8 $\mu_B$/Fe atom. From theoretical
considerations we find that the formation of a local moment on
neighboring iron sites of an AV sets in when the local Coulomb
interaction exceeds a critical value of $\sim 1$ eV in the dilute
limit. Its estimated value amounts to $\sim 2.5$ eV and implies an
upper bound of $\sim 2$ eV for the Coulomb repulsion at Fe sites
beyond the first neighbor-shell of an AV. Electronic correlations
are thus moderate/weak in doped La-1111. The strongly enhanced
spin susceptibility is responsible for the Pauli limiting behavior
of the superconductivity that we observe in As-deficient ${\rm
LaO_{0.9}F_{0.1}FeAs_{1-\delta}}$. In contrast, no Pauli limiting
behavior is found for the optimally doped, As-stoichiometric ${\rm
LaO_{0.9}F_{0.1}FeAs}$ superconductor in accord with its low spin
susceptibility.}
\end{abstract}
 \maketitle

\section{Introduction}
Since the discovery of superconductivity in the Fe-pnictides
\cite{Kami08} great efforts have been made to understand the
unusual physical properties of these systems. Most of their parent
compounds are viewed as itinerant antiferromagnets with a spin
density wave (SDW),\cite{Cruz08,Rott08} although the strength of
correlation effects is still under
debate.\cite{Kut10,Hozoy09,Yang09,Drechs09} Superconductivity appears by
doping, if the antiferromagnetic (AFM) ordering is suppressed. On
the other hand, upper critical field measurements at high magnetic
fields have shown that many of the iron-based superconductors are
limited by the Pauli paramagnetism.\cite{Fuchs08,Fuchs09} This
limitation should be related to a large paramagnetic spin
susceptibility of the conducting electrons in the normal state
which mediates the pair-breaking of singlet
Cooper-pairs.\cite{Fuchs09} So far, to the best of our knowledge,
the expected relationship between the Pauli limiting behavior and
an enhanced spin susceptibility in the normal state has not
yet been confirmed experimentally for the Fe-pnictide
superconductors\cite{remark}

A strongly enhanced susceptibility $\chi_s(q=0)$ would put these
systems closer to a ferromagnetic (FM) instability, and it
requires a sizable Stoner factor. For example, according to recent
investigations \cite{Sing08} the La-1111 parent compound is
already close to such a  ferromagnetic
instability which competes with the predominant Fermi surface
nesting driven antiferromagnetic instability.
The vicinity of Fe-pnictides and related systems to several
competing magnetically ordered and superconducting phases
seems to be a generic, but not yet well studied, feature.
In this respect, even a
relatively small increase of the Stoner factor will result in a
sizable enhancement of the paramagnetic susceptibility. It was
also pointed out\cite{Fuchs09} that the local magnetic field can
be enhanced by strongly paramagnetic (PM) centers, AFM or
ferromagnetic secondary phases coexisting with the superconducting
main phase. For instance, the AFM compound ${\rm Fe_2As}$ or
others might in high fields be converted into a highly polarized
magnetic state.\cite{Fuchs09} A general theoretical
consideration of possible underlying microscopic mechanisms
responsible for the enhanced susceptibility and its relations to
the Pauli limiting behavior is still lacking. The La-1111 pnictide
is a good model system for such investigations. Recently, Pauli
limiting behavior has been found there for optimally doped,
polycrystalline ${\rm LaO_{0.9}F_{0.1}FeAs_{1-\delta}}$ samples
with As-vacancies (AV) in the concentration  range of $\delta \sim
0.05-0.1$.\cite{Fuchs08,Fuchs09,Hammer10} In contrast,
As-stoichiometric "clean" ${\rm LaO_{0.9}F_{0.1}FeAs}$ samples with
nearly the same F-doping level do not show any Pauli limiting
behavior.\cite{Hunte08}

The first indication for an enhanced paramagnetism in As-deficient
samples (compared with As-stoichiometric reference samples) 
came from 
a strong
exponential relaxation of the muon spin polarization observed in
$\mu$SR measurements.\cite{Fuchs09} The authors of Refs.\
\onlinecite{Fuchs08,Fuchs09} supposed that disorder in the As-deficient
sample gives rise to the formation of dilute quasi-static
paramagnetic spin clusters of unknown origin. Here we will
demonstrate by a comparative analysis of the static susceptibility
data and the $^{75}$As-nuclear quadrupole resonance (NQR) spectra
together with the nuclear spin-lattice
relaxation rate $1/T_1T$ of As-deficient samples that the very
vicinity of an AV provides a direct candidate for such paramagnetic
centers.

Our paper is organized as follows. In Sect.\ II we consider briefly
what is known about the concentration of As-vacancies and how the actual 
concentration can be refined using NQR spectroscopy. Sect.\  III concerns 
with the magnetic susceptibility. In the first subsection III.A it is 
explained
how the effect of ferromagnetic inclusions is eliminated to get the 
intrinsic susceptibility 
analyzed in the second subsection III.B.
Then  we consider theoretical aspects of localized states and local
magnetic moments in the frame of Wolff's approach to local moments
in a nonmagnetic host.\cite{Wolf61} In Sect.\ IV we apply this gained insight to 
estimate the effective Coulomb repulsion for a localized state
derived from the As-vacancy and arrive at an upper bound for the Hubbard 
$U$  on the Fe-sites which bear the superconductivity. In Sect.\ V we
discuss briefly how the local magnetic moments affect the NMR data.
Local moment related aspects of Pauli limited superconductivity are 
considered in Sect.\ VI. In Sect. VII we briefly mention a similar
situation in Sn-flux grown Ba-122 superconductors.
Finally we end up with a conclusion containing
the gained insight and perspectives for future work.

\section{The 
concentration of As-vacancies and the NQR
spectra}\label{sect2}
\begin{figure}[t]
\includegraphics[width=8cm,angle=0]{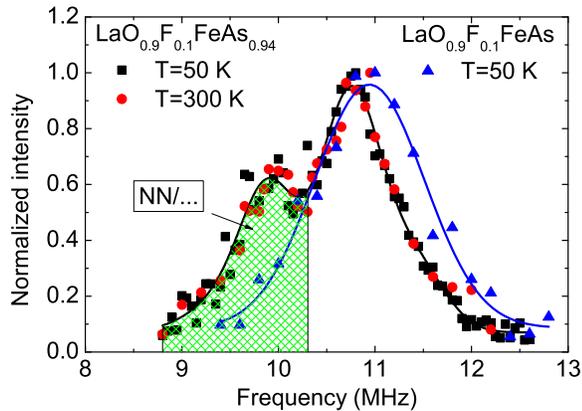}
\caption{(Color online) 
$^{75}$As NQR spectrum at $T=$50~K, together with the spectrum of a 
 reference sample \cite{Lang10b}, and 
 room-temperature 
 measurements \cite{Lang10a}. Black line: a typical fit according to an 
 As-vacancy 
 arrangement shown in Fig.\ \ref{f.2}. Blue line: broad single-peak fit
 of the NQR-spectrum of  a reference sample. 
 NN/...\   and the shadded green area
 indicate schematically
 the low-frequency
 spectral weight below about 10.3~MHz 
 suggested for the  NN and NNN shells around an AV as shown in
 Fig.\ \ref{f.2}.
 } 
\label{f.1}
\end{figure}
Polycrystalline ${\rm LaO_{0.9}F_{0.1}FeAs_{1-\delta}}$ samples
were prepared from pure components using a two-step solid state
reaction method \cite{Kond09}. ${\rm
LaO_{0.9}F_{0.1}FeAs_{1-\delta}}$ samples were obtained by
wrapping the samples in a Ta foil during the annealing procedure.
\cite{Fuchs08,Fuchs09} According to  energy dispersive x-ray
(EDX) analysis, an As/Fe ratio of about 1.0 was found in the
reference sample annealed without a Ta foil and of about 0.9 to
0.95 in the As-deficient samples. At first glance one might expect
an AV-gradient within the ${\rm
LaO_{0.9}F_{0.1}FeAs_{1-\delta}}$ samples because the vacancies
start to be formed at the surface of the
samples. Nevertheless, the sharp
superconducting transition width comparable with that of 
As-stoichiometric
reference samples \cite{Fuchs09} and the surprising temperature
dependence of the nuclear spin-lattice relaxation rate $T_1^{-1}
\sim T^5$ (the reference samples show $T_1^{-1} \sim  T^3$
instead\cite{Hammer10,Grafe08})
indicate a homogeneous AV-distribution 
within the sample. Additionally, 
the static susceptibilities $\chi_p(H,T)$ of the
As-deficient samples 
in both the bulk and the surface parts
were the same within
the error bars of our measurements.
\begin{figure}[b]
\includegraphics[width=8.5cm,angle=0]{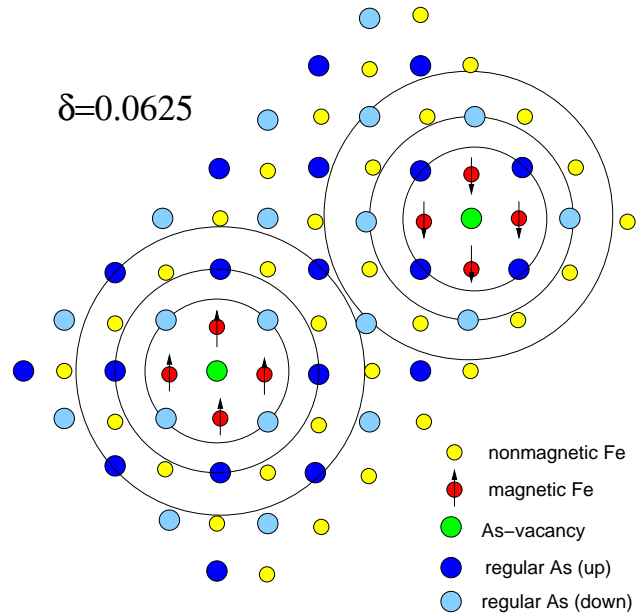}
\caption{(Color online) Schematic structure of an FeAs-block with
two neighboring
As vacancies (AV) at a  concentration of $\delta =
0.0625$, i.e.\ in the range suggested by the NQR-data shown in 
Fig. \ref{f.1} for a
typical As-deficient sample. 
The low-frequency spectral weight  below 10.3~MHz 
is attributed to NN and NNN-shells
around an As-vacancy (AV) whereas the high-frequency part
is attributed to to more distant As sites.
Notice the four Fe-sites surrounding an AV involved in a local moment. The
local magnetic moments are depicted for the simplest case where only
the first neighbor shell around an AV is affected (see
also text).} 
\label{f.2}
\end{figure}

A local characterization was performed using 
the NQR technique. 
The $^{75}$As NQR spectrum of an As-deficient sample
at $T=$50~K is shown in Fig.~\ref{f.1}  
together with that of a reference sample \cite{Lang10b}.
Similar 
room-temperature data
\cite{Lang10a} 
point to a negligible
temperature-dependence. For As nuclei (spin $I=$3/2), the measured 
frequencies obey $\nu_Q \propto Q V_{zz} \sqrt{1+\eta^2/3}$ with $Q$ the 
electric quadrupole moment, $V_{zz}$ the largest eigenvalue of the electric 
field gradient tensor, and $\eta$ the asymmetry of the latter. 
Whereas the reference samples feature 
a broad smooth distribution 
 of charge environments,
the samples with AVs show several components, as could be expected 
from varying distances between As nuclei and charged AVs
(see also Section III.C). Since 
the high-frequency spectral weight is at frequencies rather similar to 
those of the reference samples, it is likely associated to the As nuclei 
far away from an AV. The low-frequency spectral weight would 
accordingly correspond to sites closer to an AV (see Fig.\ 2).
The data was fitted with up to four components (two of them for the 
low-frequency weight) and assuming that the low-frequency weight corresponds 
to nearest neighbors (NN) or to both NN and next-nearest neighbors (NNN). In 
the latter case, two components of equal areas were used, reflecting the 
assumption that electrostatic repulsion separates the vacancies enough that 
each of them features 4 NN and 4 NNN (see Fig.~\ref{f.2}). Since 
the ratio of low-frequency to high-frequency weight is then 
$4 \delta / (1-5\delta)$ (NN) or $8 \delta / (1-9\delta)$ (NN+NNN), 
this leads to $\delta = 0.06(2)$, with the error bar accounting for 
different fitting procedures. 
Therefore, the NQR measurements indicate that 
the AVs are  almost homogeneously
distributed within the sample volume in the amount as expected from EDX 
measurements. Future study at other compositions will aim at refining 
this approach, including also the antiferromagnetic parent compound 
and LiFeAs derived As-deficient samples
without additional disorder caused by  F-dopants.
As a consequence, the widths of the $^{75}$As-NQR line in both parent 
compounds are very small $\sim 0.1-0.2$ MHz \cite{Morozov10,Lang10b}
in sharp contrast to our F-doped  samples. 

\section{Static susceptibility}\label{sect3}

\begin{figure}[t!]
\includegraphics[width=8.5cm,angle=0]{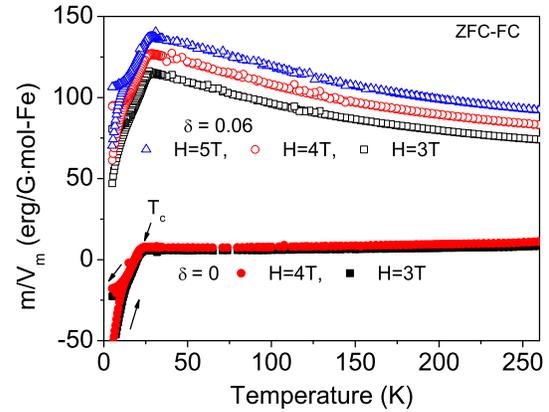}
\caption{(color online) Temperature dependence of the
magnetization after zero-field cooling at different magnetic
fields for ${\rm LaO_{0.9}F_{0.1}FeAs_{1-\delta}}$. } \label{f.3}
\end{figure}

In view of the rather specific field-  and \emph{T}-dependencies
of the static magnetization 
of the As-deficient samples
and its large magnitude (see Fig.\ \ref{f.3}),  
a more sophisticated analysis of the magnetization is required. 
The magnetization
consists of three main contributions arising (i) from a
ferromagnetic (FM) contribution which we attribute to
Fe-inclusions (see Section III.A), (ii) from localized magnetic
moments in the very vicinity of a given AV, and (iii) from the
\emph{T}-dependent susceptibility of the Fermi sea of itinerant
conduction electrons in which the AV are embedded. The last two
contributions are strongly related with each other and can be
understood in the framework of Wolff's theoretical approach to
impurity-effects \cite{Wolf61} (see Section III.C).

\subsection{Iron inclusions}\label{sect3.1}

\begin{figure}[t!]
\includegraphics[width=8.99cm,angle=0]{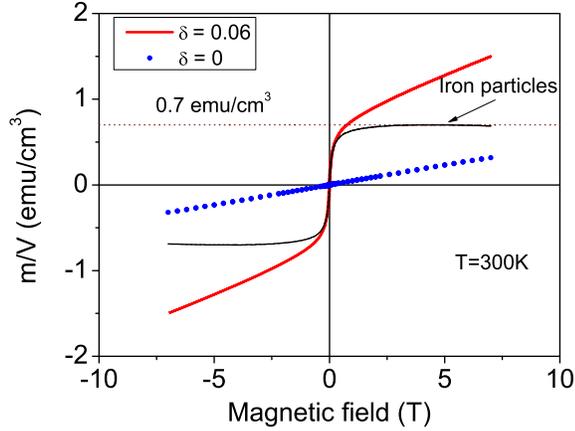}
\caption{(color online) Field dependence of the magnetization of
${\rm LaO_{0.9}F_{0.1}FeAs_{1-\delta}}$ at 300 K. } \label{f.4}
\end{figure}

\begin{figure}[b!]
\includegraphics[width=9.5cm,angle=0]{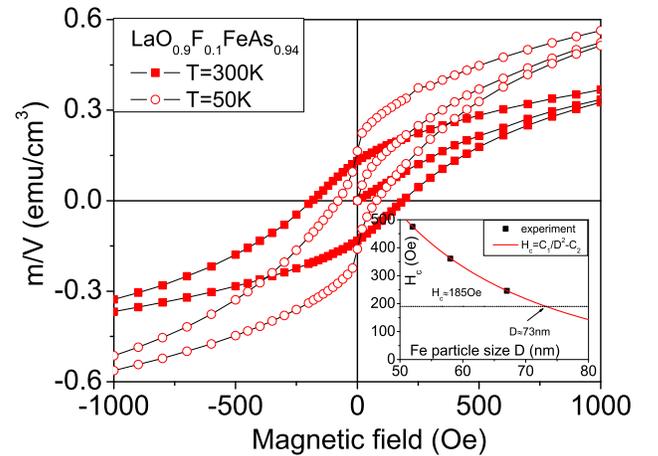}
\caption{(color online) Field dependence of the magnetization of
${\rm LaO_{0.9}F_{0.1}FeAs_{0.94}}. $ Inset: Coercive field $H_{cor}$
of iron particles vs. particle size (these experimental data 
are
taken from
Ref. \cite{Dumas07}). The intercept between the fit curve and the
horizontal line at 185 Oe gives the minimum size of the iron
particles in the $\rm LaO_{0.9}F_{0.1}FeAs_{0.94}$ sample. }
\label{f.5}
\end{figure}

In Fig. \ref{f.4}, the field dependence of the volume magnetization of two
typical ${\rm LaO_{0.9}F_{0.1}FeAs}$ and  ${\rm
LaO_{0.9}F_{0.1}FeAs_{1-\delta}}$ 
samples are compared at \emph{T}
= 300 K. (The magnetization measurements were performed in a
Quantum Design DC SQUID.) It is seen that ${\rm
LaO_{0.9}F_{0.1}FeAs}$ has a small paramagnetic magnetization with
almost linear field dependence. In sharp contrast, ${\rm
LaO_{0.9}F_{0.1}FeAs_{1-\delta}}$ exhibits a considerably higher
magnetization. Its field dependence and the presence of the
magnetic hysteresis (see Fig. \ref{f.5}) is an indication of
ferromagnetic (FM) behavior. At high magnetic fields, the
magnetization of a FM material saturates and only a linear
paramagnetic contribution remains (see Fig. \ref{f.4}). We
suppose that the FM contribution stems from inclusions of pure Fe
particles in the As deficient samples because of (i) a
high value of
the coercitivity field ($H_{\rm cor}$) and (ii) a
rather high
Curie-temperature of the FM ordering $T_C > 360$K. These large Fe
particles are formed during the heat treatment in contact with a
Ta-foil, since no ferromagnetic inclusions have been observed for
the As-stoichiometric reference compound obtained in the same manner despite
the final heating in a Ta-foil which produces the AV. In fact, a
possible scenario might be: in many Fe-pnictides there is a small
amount of FeAs$_2$, FeAs,\cite{Baker08} or  other
antiferromagnetic inclusions which become ferromagnetic, and probably
are pure Fe inclusions, if a predominant part of As is extracted from
them, too. Thus, in the extraction process As is taken from two
kind of regions: from the pristine regions and from those with
inclusions.

The field $H_{\rm cor}$ is related to the particle size of the
adopted Fe-inclusions. We obtained for the ${\rm
LaO_{0.9}F_{0.1}FeAs_{1-\delta}}$ samples a coercivity field of
$H_{\rm cor} \sim ~ 185$ Oe at 300 K (see Fig. \ref{f.5}). Using
experimental data for the dependence of $H_{\rm cor}$ on the size
of the Fe-particles \cite{Dumas07} we estimated for them a size of
about 75 nm (see inset in Fig. \ref{f.5}). It is known that the
saturation induction of FM materials is independent of the
particle size. This allows us to estimate the fraction of the Fe
particles in the samples from the magnetization value of ${\rm
LaO_{0.9}F_{0.1}FeAs_{1-\delta}}$. At high magnetic fields, the
paramagnetic contribution of the sample can be subtracted from the
total magnetization and the field-independent saturation
magnetization is proportional to the Fe fraction in the sample
(Fig. \ref{f.4}). At \emph{T}= 300 K, the saturation
magnetization of iron $M_{\rm s,Fe} \approx 1.7\cdot 10^3$
emu/cm$^3$ (see Ref.\onlinecite{Dumas07}), whereas the ${\rm
LaO_{0.9}F_{0.1}FeAs_{1-\delta}}$ samples have at 300 K a saturation
magnetization of about $M_{\rm s,def} \approx$ 0.7emu/cm$^3$.
Then, with the unit cell volume 
of the As-deficient sample \cite{Fuchs09} 
$v_{\rm unit} \approx 0.1415nm^3$
(i.e.\ with two Fe atoms per cell) and the sample filling factor of 
$n^{}_{\rm
fill} \approx 0.64$, we
estimate for the ratio between Fe atoms in the few FM inclusions
and the regular Fe atoms in the As-deficient samples:
\begin{equation}
n_{\rm Fe} =\frac{v_{\rm unit}N_a}{2n_{\rm
fill}}\cdot\frac{\rho_{\rm Fe}M_{\rm s,def}}{m_{\rm Fe}M_{\rm s,
Fe}} \approx 3.9\cdot 10^{-3} ,
\end{equation}
where $N^{}_a$ denotes the Avogadro constant, $m^{}_{\rm Fe}$ is
 the atomic mass of iron, and 
$\rho^{}_{\rm Fe}$ its density.
Thus, we estimate a small atomic fraction of Fe atoms of
about 0.4$\%$ residing within these ferromagnetic inclusions.
Hence, such a small amount of iron has no influence on the
estimated effective Fe excess due to the AV but it nevertheless
strongly affects the magnetization curves of the As-deficient
samples. Therefore
this inclusion contribution should be subtracted
to get the information concerning the static
susceptibility .

\subsection{\emph{T}-dependence of the paramagnetic
susceptibility}\label{sect3.2}

As is shown in Fig. \ref{f.4}, the field dependence of the total
magnetization becomes linear for magnetic fields exceeding 2T.
Thus, we conclude that the Fe inclusions completely saturate at
these fields. To subtract the iron particles contribution, the
\emph{T}-dependence of the magnetic moment of an As-deficient sample
after zero-field cooling (ZFC) and at field cooling (FC) was
measured at several fields above 2T (see Fig.\ \ref{f.3} ). 
For comparison,
the \emph{T}-dependence of the magnetic moment of a reference
sample ${\rm LaO_{0.9}F_{0.1}FeAs}$ is also shown at different
fields. To get the intrinsic paramagnetic susceptibility of ${\rm
LaO_{0.9}F_{0.1}FeAs_{1-\delta}}$, we subtracted the magnetic
moment $m_1(T)$ measured at the field $H_1$ from the moment
$m_2(T)$ measured at the field $H_2$ and divided this difference
by the corresponding field difference:
\begin{equation}\label{3.1}
 \frac{\Delta m^p}{V_m\Delta
 H}=\frac{m^{}_2(T)-m^{}_1(T)}{V_m(H_2-H_1)}=\frac{m^p_2(T)-m^p_1(T)}{V_m(H_2-H_1)}\approx
 \chi^{}_p(T).
\end{equation}
Only paramagnetic moments ($m^p$) can contribute to  $\Delta m =
m_2(T) - m_1(T)$ at high fields because the FM contribution is
already saturated. At high temperatures, Eq. (\ref{3.1}) gives the
static paramagnetic susceptibility $\chi_p$ in [emu/mol-Fe]. Here
$V_m$ is the number of moles per Fe atom. The same procedure was
done also for ${\rm LaO_{0.9}F_{0.1}FeAs}$ as a reference sample.

\begin{figure}[t]
\hspace{-1.5cm}
\includegraphics[width=9.5cm,angle=0]{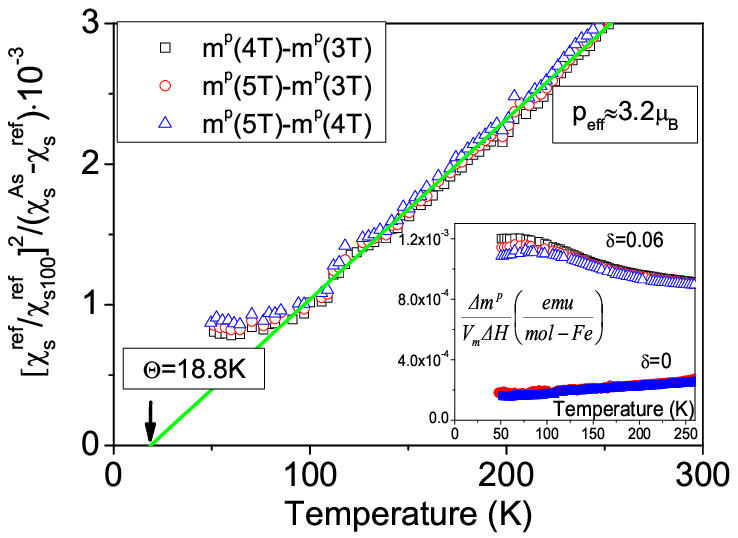}
\caption{(Color online) The $T$-dependence of the ratio $
\left[\frac{\chi^{\rm ref}_{s}(T)}{\chi^{\rm ref}_{s}(T^{}_{\rm
ref})}\right]^2\cdot \frac{1}{\chi^{\rm As}_{s}(T)-\chi^{\rm
ref}_{s}(T)}$ for ${\rm LaO_{0.9}F_{0.1}FeAs_{1-\delta}}$, where
$\chi_s^{\rm ref}$ and $\chi_s^{\rm As}$ are spin susceptibilities
of the ${\rm LaO_{0.9}F_{0.1}FeAs}$ and ${\rm
LaO_{0.9}F_{0.1}FeAs_{0.94}}$. Straight line is a fit by Eq.
(\ref{3.4}). Details of the fitting procedure are described in the
text. The ratio $\frac{\Delta m^p}{V_m\Delta H}\left(\frac{{\rm
emu}}{\rm mol-Fe}\right)$ for two samples defined by Eq.
(\ref{3.1}) is shown in the inset. The closed symbols correspond
to ${\rm LaO_{0.9}F_{0.1}FeAs}$ data for the $m^p(4{\rm
T})-m^p(3{\rm T})$ and $m^p(2{\rm T})-m^p(1{\rm T})$. }
\label{f.6}
\end{figure}

The \emph{T}-dependence of the ratios $\Delta m^p/V_m\Delta H$ for
both samples above $T_c$ is shown in the inset of Fig. \ref{f.6}. For
${\rm LaO_{0.9}F_{0.1}FeAs}$, the susceptibility $\Delta
m^p/V_m\Delta H$ linearly increases with the temperature which is
typical for this compound where the AFM spin ordering has been
completely suppressed. The absolute value of the susceptibility of
the reference samples is similar to that reported by Klingeler et
al. \cite{Kling10} In contrast, for
${\rm LaO_{0.9}F_{0.1}FeAs_{1-\delta}}$ the paramagnetic susceptibility
is considerably
higher and exhibits an unusual T-dependence.

First, we suppose that AV can induce local moments.
These moments, for example, might occur on a microscopic level
just in the vicinity of AV. \cite{Fuchs09} Possible mechanisms of
local moment formation are discussed in Section III.C. If so, the
total number of local magnetic moments in the sample is
proportional to the number of AVs $\delta$. These local
moments lead to a Curie-Weiss-like contribution at high
temperatures \cite{Clog62} (see Figs. \ref{f.3} and \ref{f.6}),
\begin{equation}\label{3.2}
\chi^{\rm As}_{\rm Curie}(T) \approx \frac{C}{T-\Theta}
\end{equation}
with
\begin{equation}\label{3.3}
C = N_a \delta\frac{\left( p^{}_{\rm eff}\mu_{\rm B}\right)^2}{3k^{}_{\rm B}}
\end{equation}
where $\Theta$ is the Curie temperature characterizing the
effective strength of magnetic interactions averaged over the
whole sample. Its sign reflects the type of this exchange
interaction (FM or AFM). In Eq.\ (\ref{3.3})  
\emph{C} denotes the Curie constant, $p_{\rm
eff}=\textrm{g}[J(J+1)]^{0.5}$, $\mu_{\rm B}$ is the Bohr magneton, g is
the Land\'e factor, \emph{J} stands for the total electronic angular
moment and $k_{\rm B}$ is the Boltzmann constant.

Second, these moments are formed in a metallic host with a
\emph{T}-dependent susceptibility. We suppose that this 'bare'
host susceptibility of the As-deficient sample is the same as the
paramagnetic susceptibility $\chi^{\rm ref}_p(T)$ of the reference
sample. For local moments in a matrix with a
\emph{T}-dependent susceptibility, the experimental data can be
analyzed using a method similar to that described in Ref.\ \onlinecite{Clog62}.
Therefore, we suppose that the value of the local moment $p_{\rm
eff}$ is proportional to the spin susceptibility of the matrix:
\begin{equation}
p^{}_{\rm eff}(T) = \frac{p^{}_{\rm eff}(T^{}_{\rm
ref})}{\chi^{\rm ref}_s(T^{}_{\rm ref})}\cdot\chi^{\rm ref}_s(T),
\end{equation}
where $T_{\rm ref}$ is some reference temperature. In our case we
have used 
$T_{\rm ref} = 100$~K as the lowest temperature at which the local
paramagnetic susceptibility can be reasonably described by a Curie-like
function [see Eq.\ (\ref{3.4}) below and  Fig.\ \ref{f.6}]. 

The spin susceptibility $\chi^{\rm ref}_s$ can differ from the
paramagnetic susceptibility $\chi^{\rm ref}_p$ due to additional
contributions $\chi_{\rm chem}$ from various types of diamagnetism
and from the Van Vleck paramagnetism. According to Ref.\
\onlinecite{Sing08} the estimated (bare) Pauli susceptibility of
LaOFeAs from the calculated electronic density of states 
(DOS) at the Fermi level, $N(\varepsilon_F)$, 
amounts $\chi_0 \approx 8.5\cdot 10^{-5}$ emu/mol-Fe.
But the actual experimental value \cite{Kling10} amounts
$\chi_s = \chi_0\cdot(1-I)^{-1}\geq 4\chi_0$,
where $I=JN(\varepsilon_F)$ 
i.e.\ it is
significantly enhanced due to the presence of a sizable Stoner 
factor \cite{Sing08} , where $J \sim $0.7-0.8 eV (see also IV). It is 
known \cite{Sing08,Lars09,Boer08}
that $N(\varepsilon_F)$ slightly decreases with F doping. 
This leads to a decrease of both $\chi_0$ and $N(\varepsilon_F)J$. 
From the analysis of the experimental data reported in
Ref.\ \onlinecite{Kling10} one can see that at high temperature
(above $T_{\rm N}$ of the AFM ordering) the value of the susceptibility
reduces only on ~15\% at the doping level of about 
$\sim 0.1.$ Therefore, one  expects that  $\chi^{\rm ref}_s$ of the
reference sample is also strongly enhanced due to a Stoner factor with the 
corresponding bare spin susceptibility $\chi_0^{\rm ref} \approx 
 7\cdot 10^{-5}$~emu/mol-Fe. From this point of view we expect that the spin
susceptibility $\chi^{\rm ref}_s$ has a dominant contribution to the measured
paramagnetic susceptibility $\chi^{\rm ref}_p$ 
of the reference samples. Hence we suppose that 
$\chi^{\rm ref}_s \approx \chi^{\rm ref}_p$.

Finally, according to Ref. \onlinecite{Clog62},
 the effective susceptibility of a metal with additional
extrinsic local moments can be written at high temperatures $T>
\Theta$ in the form:
\begin{equation}\label{3.4}
 \chi^{\rm As}_{s}\approx\chi^{\rm ref}_{s}(T)\cdot\left[ 1 +
\frac{C(T^{}_{\rm ref})}{(T-\Theta)}\cdot\frac{\chi^{\rm
ref}_s(T)}{\chi^{{\rm ref}^2}_s(T_{\rm ref})}\right],
\end{equation}
Our experimental data can be reasonably well described by Eq.
(\ref{3.4}) for $T >80$~K. The ratio
$$
\left[\frac{\chi^{\rm ref}_{s}(T)}{\chi^{\rm ref}_{s}(T^{}_{\rm
ref})}\right]^2\cdot \frac{1}{\chi^{\rm As}_{s}(T)-\chi^{\rm
ref}_{s}(T)}
$$
is plotted in Fig. \ref{f.6} where the susceptibility $\chi_s
\approx \chi_p$ is defined by Eq. (\ref{3.1}). The fit by Eq.
(\ref{3.4}) (the straight line) yields $p_{\rm eff} \approx 3.2$
which corresponds to an AV concentration of $\delta \approx 0.06$
and a Curie temperature $\Theta \approx 18.8$ K pointing to dominant
FM correlations
between the Fe electrons.

\subsection{The formation of magnetic moments}\label{sect3.3}

It is well-known that nonmagnetic impurities like Zn induce local
moments on neighboring Cu atoms in ${\rm
YBa_2Cu_3O_{7-\delta}}$.\cite{Mahaj94,Allo09} This effect is not
too puzzling because formally isovalent Zn$^{2+}$ impurities
substitute for magnetic Cu$^{2+}$ ions in its CuO$_2$ layers and
thus break the Zhang-Rice singlet states. In case of
 ${\rm LaO_{0.9}F_{0.1}FeAs_{1-\delta}}$
 the origin of the local moments induced by As
vacancies is less obvious, but eventually the moment formation can
be understood if a strong enough \emph{d-p }hybridization between
the As 4\emph{p} and Fe 3\emph{d} orbitals \cite{Garc08} is taken
into account. An AV removes the covalent bonds with 3\emph{d}
orbitals from four adjacent Fe ions, so the actual defect in the
quasi 2D Fe-As layer is a [${\rm V_{As}Fe_{4}}$] complex with
dangling \emph{d-p} bonds (cf. Fig. \ref{f.2}). Several effects
are related to the 
formation of this complex. First, the charge
transfer from Fe ions to the empty As site results in a local
enhancement of the effective charge around Fe ions. The analysis
of the reflectivity supports this assumption.\cite{Fuchs09, Drechs09}
Second, due to the same charge redistribution the distance of As
atoms from the basal Fe plane in ${\rm
LaO_{0.9}F_{0.1}FeAs_{1-\delta}}$ increases as compared to ${\rm
LaO_{0.9}F_{0.1}FeAs}$.\cite{Fuchs09} But the most interesting
effect is the possibility of a 
formation of localized states and related
non-compensated magnetic
moments around an AV. A more detailed microscopic study will be reported
elsewhere \cite{Kikoin11}. Here we restrict ourselves mainly to qualitative
aspects.

The basic conditions for formation of a non-zero magnetic moment
around a non-magnetic impurity in a
paramagnetic metallic host have
been formulated in Refs. \onlinecite{Wolf61,Mills67}. In simple
terms, a magnetic moment may be formed provided the onsite Coulomb
repulsion \emph{U} of two electrons exceeds some critical value
$U_c$ estimated as
\begin{equation}\label{3.5}
\frac{U^{}_c}{E^{}_B}\approx \frac{1}{2} +\frac{(\varepsilon_F -
E_0)^2}{2\Delta^2},
\end{equation}
where $E_B$ is the effective bandwidth, $\varepsilon_F$  is the
Fermi energy, $E_0$ and $\Delta$ are the position and the width of
the resonance, respectively, created by the defect in the band. In
our case \emph{U} is the Coulomb repulsion of two electrons
occupying the hybridized \emph{d-p} orbitals in the [${\rm
V_{As}Fe_{4}}$] complex and the carriers occupy hole pocket around
$\Gamma$ point in the Brillouin zone. The ratio (\ref{3.5}) may be
large enough because each of the 4 Fe ions donates part of its
Hubbard repulsion $U_0$ to the repulsive interaction between
the\emph{ d-p} "molecular orbitals" in the As-vacancy complex. In
accordance with this spin-dependent scattering mechanism, a single
AV causes the formation of a magnetic moment shared between the
adjacent Fe ions. The net magnetic moment associated with a [${\rm
V_{As}Fe_{4}}$] complex defect amounts about $p_{\rm eff} \approx
3.2$ according to above estimates. This value corresponds to
$\approx 0.8 \mu_B$/Fe atom, if the magnetic moments occur in the
first neighbor shell, only. However, we cannot exclude that the
second shell is affected as well. In such a case one is left with
two options: parallel or antiparallel spin orientations. Similar
results have been obtained in Ref. \onlinecite{Lee08} using LSDA
calculation (i.e. ignoring the local Coulomb repulsion \emph{U})
for ${\rm FeSe_{0.875}/FeTe_{0.875}}$ superstructures, i.e. with a
twice as large nominal concentration of vacancies and a stronger
mutual influence as compared with our title compound and
As-vacancies. In the former case, the magnitude of the
anti-parallel oriented moments at the second neighbor shell was
about 1/4 - 1/3 to that of the first neighbor shell. In other
words, then a relatively large, but far from saturation
($2\mu_B$), moment of about 1.067 - 1.2 $\mu_B$ would reside at
any Fe site within the first neighbor shell.

A detailed analysis of various spectroscopies might be helpful to
elucidate the corresponding local magnetic structure. The
relatively large value of the local magnetic moments estimated
above should be compared with the regular magnetic moments $\sim
0.4\mu_B$ in the magnetically ordered parent compound
LaOFeAs.\cite{Cruz08} Thus, the magnetic moment induced by an
As-vacancy is about 2 to 3 times larger than the experimental
value observed in stoichiometric LaOFeAs phase but less than the
theoretical value ($\sim 2.3\mu_B$) incorrectly predicted by the
L(S)DA and other modern band structure calculations.\cite{Cruz08}
The resolution of this puzzle is one of the central problems for a
future microscopic theory of iron pnictides.

It should be stressed that in accordance with Wolff's approach 
\cite{Wolf61} the
[${\rm V_{As}Fe_{4}}$] complexes are isoelectronic defects and
thus do not affect the carrier concentration. The magnetic
moment arises due to the spin dependent local density of
electronic states in the hole-band near the Fermi level. This
defect-related structure in the density of states consists of two
nearly Lorentzian peaks centered below $\varepsilon_F$ (majority
spin peak) and above $\varepsilon_F$ (minority spin peak). Due to
the symmetry related selection rules, the influence of magnetic
scattering on the behavior of the electrons in the electronic band
is reduced by a factor of 
$\sim |{\bf q}|/|{\bf G}|$, where $\bf q$
is the deviation of the scattering vector from the nesting vector
$\bf G$. The presence of such peaks may be detected
experimentally.

It is important to realize that the strongly anisotropic structure
of LaOFeAs prevents clustering of AV since these
clusters would induce a large local charge in the Fe plane and
strongly increase the potential energy of the lattice. Thus, the
lattice anisotropy tends to make the distribution of the AV
uniform. This explains why the AVs are relatively homogeneously
distributed in our
As-deficient samples in spite of the somewhat
uncontrollable method of the AV formation.
In turn, the charged AVs  seem to make
the F-distribution more homogeneous. This effect can
qualitatively explain why As-deficient samples exhibits 
narrower NQR
peaks (Fig.\ref{f.1}) compared to the reference sample. Moreover,
since the size of these magnetic defects is small in comparison
with the superconductor coherence length, the former cannot
essentially reduce the superconducting volume fraction.

Thus, the significant enhancement of the spin susceptibility
$\chi_s^{\rm As}$ in the As-deficient samples compared to
$\chi_s^{\rm ref}$ (Fig. \ref{f.6}, inset) may be ascribed to an
additional contribution of AV related magnetic defects to the
magnetic response. Then in accordance with the predictions of the
Wolff model, the susceptibility of a metal containing few
impurities with a short-range scattering potential and a strong
enough AV related local Coulomb repulsion factor $U_v$ has
the form: \cite{Mills67}
\begin{equation}\label{3.6}
\chi^{\rm As}_s =\chi^{\rm ref}_{s}\left[1 + \frac{2\delta U_v J
\chi_s^{\rm ref}/R}{1- U_v J_l \chi_l/R}\right]
\end{equation}
where $\delta \approx 0.06$ is the concentration of the local
defects as determined from the 
EDX and the NQR-data mentioned above, $\chi_l =
\langle S^+S^-\rangle_l$  is the local transverse susceptibility
of a magnetic defect, $R = 4(\rm{g}\cdot \mu_B/2)^2N_a/k_B=1.5
{\rm emu\cdot K/mol} \approx 1.29\cdot 10^{-4} \rm{emu\cdot
eV/mol}$ and \emph{J}=0.7eV (is a typical value for 3\emph{d}
electrons in Fe). Since \emph{U} is supposed to be large enough,
the observed enhancement of the spin susceptibility and its
$\delta$-dependence for our
As-deficient samples can be understood at least qualitatively.
At high temperatures Eq.\ (\ref{3.6}) yields a Curie-Weiss-like
behavior similar as Eq.\ (\ref{3.4}). Comparing these equations we see that
the temperature $\Theta \approx 18.8$ K characterizes either some
FM correlations between itinerant electrons scattered on the AV or
short-range correlations between the localized electrons.

It follows from Eq. (\ref{3.6}) that the susceptibility
$\chi_s^{\rm As}$ increases with increasing AV concentration
$\delta$. Preliminary data obtained for As-deficient samples with
different $\delta$ confirm such a behavior. At high AV
concentrations, a deviation 
from the linear dependency 
of $\chi_s^{\rm As}$ on $\delta$ is expected. Therefore, further
analysis is required to understand the range of applicability of
Eq. (\ref{3.6}).

\section{Estimation of the local Coulomb repulsion }

\begin{figure}[t!]
\includegraphics[width=9.5cm,angle=0]{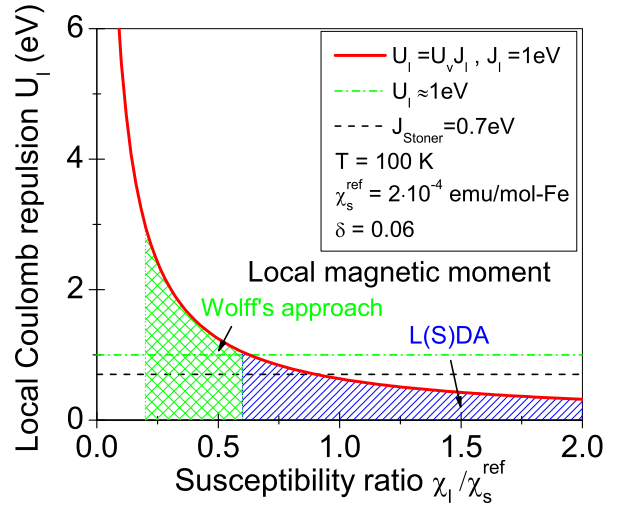}
\caption{(Color online) The local Coulomb repulsion for a
localized state created by an As vacancy according to Eq.
(\ref{3.6}). The green line denotes the critical strength of $U_l$
where the formation of a local moment at iron sites surrounding an
As-vacancy sets in according with the Wolff's model. Noteworthy,
we admit that local moments can exist also below $U_l = 1$eV.
According to L(S)DA calculations reported in Ref.
\onlinecite{Lee08} the formation of the local moments can be
expected also in a weakly correlated situation since there is no
correlation in the L(S)DA approach.  }
 \label{f.7}
\end{figure}

In general, the parameter $U_vJ_l$ and the local susceptibility
$\chi_l$ are related to each other and should be chosen in a
self-consistent way. This is a rather complicated theoretical problem. Its
solution would be of considerable interest for future
investigations. Here, we will demonstrate that from the present
simple analysis of the experimental data (see Fig.\ \ref{f.6}), 
reasonable values of these parameters can be estimated. The
relation between the local Coulomb repulsion $U_l = U_v J_l$ for a
localized state created by an As vacancy and the local transverse
susceptibility $\chi_l$ normalized per $\chi_s^{\rm ref}$ is shown
in Fig.\ \ref{f.7}. 
The green line is obtained from Eq.\ (\ref{3.5})
assuming that the position of the resonance $E_0$ is very close to
$\varepsilon_F$ and the effective bandwidth according to band
structure calculations \cite{Sing08,Vild08} amounts about $E_B
\sim 2$eV. This line denotes the critical strength of $U_l \sim
1$eV where the formation of a local moment at neighboring iron
sites of an AV sets in. In principle, measuring $\chi_l$, the
value of $U_l$ could be determined from Eq.\ (\ref{3.6}). This
would be of considerable interest since this way some new insight
into the strength of correlation effects in Fe-pnictides might be
provided. The latter is still under debate and various theoretical
estimates scatter in between 1 eV and 5 eV
\cite{Kut10,Yang09,Drechs09} although the majority of the
community supports a weak or intermediate coupling scenario.
(Thereby we assumed that $U_l$ yields an upper bound for $U_d
\approx 2$ eV on iron). This is in accord with our finding shown
in Fig. \ref{f.6}. For example, at $T_{\rm ref}  = 100$ K the
susceptibility of the reference sample is $\chi^{\rm ref}_s \approx
2\cdot10^{-4}$ emu/mol-Fe where the ratio between the
susceptibilities of the As-deficient and the reference samples
amounts 
to 
$\chi^{\rm As}_s/\chi^{\rm ref}_s \approx 6$. 
In fact, 
in view of the reasonable description achieved by our
RPA (weak coupling) based theory, we take
the susceptibility ratio $\chi_l/\chi^{\rm ref}_s$, say in
between 0.2  and 0.3, but not below.
Then
 we estimate for $J_l \approx 1$~eV
[see Eq. (\ref{4.1}) below] $U_l$ in between 2.9 and 2 eV which
provides this way an upper bound for $U_d$. Since $U_l$ should
somewhat exceed $U_d$ due to the missing screening from the
AV, we adopt also a slightly enlarged $J_l$ as compared
with the usual Stoner value of $J \approx 0.7$ eV for Fe
\cite{Sing08} regarded as a typical value for an Fe site far from
the AV in the pnictide superconductor. Then taking 
$\chi^{\rm ref} \approx  7\cdot 10^{-5}$~emu/mol-Fe (see III.B), 
we arrive at
\begin{equation}\label{4.1}
J_l \approx J\frac{1-\chi_0^{\rm ref}/\chi_s^{\rm As}}{1-
\chi_0^{\rm ref}/\chi_s^{\rm ref}} \approx 1~ {\rm eV}.
\end{equation}
With the same screening argument as used above we then may refine
this estimate: $U_d \sim 0.8 U_l \approx$ 1.6 to 2.3 eV. This
result supports the previous estimates done in Refs.\
\onlinecite{Yang09,Drechs09} and is in clear contrast with $U_d > 5$ eV
 stated recently\cite{Kut10} based on a combined RPA and dynamical
mean-field-study. In view of the L(S)DA results of Ref.\ 
\onlinecite{Lee08} mentioned above, one has to realize that a
formation of local magnetic moments might be set in already at
much smaller values of $U_l$ or larger ratios of $\chi_l/\chi^{\rm
ref}$. This gives further support for a weak-correlation scenario
at least for the occurrence of magnetic moments. To what extent a
sufficiently enhanced spin susceptibility can be obtained by that
approach, too, remains to be seen.

\section{Local moments and NMR data reanalyzed}

\begin{figure}[t]
\includegraphics[width=9.5cm,angle=0]{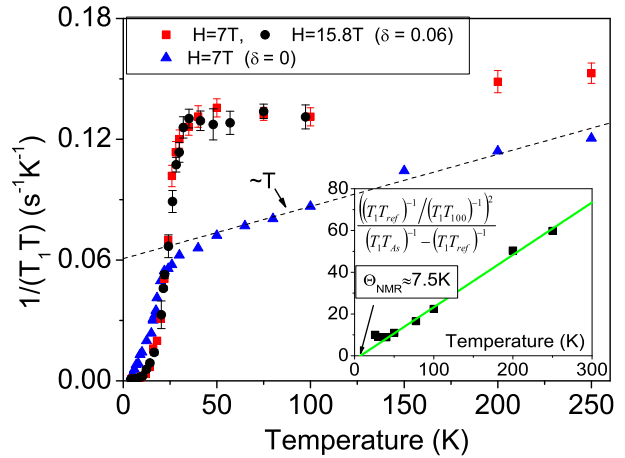}
\caption{(Color online) $^{75}$As spin-lattice relaxation rate
$1/T_1T$ versus temperature in ${\rm
LaO_{0.9}F_{0.1}FeAs_{1-\delta}}$. The data are taken from Ref.
\onlinecite{Hammer10}. The steep decrease of both curves reflects
the onset of the superconducting transition. The inset shows the
$T$-dependence of the ratio $\frac{\left((T_1T_{\rm ref})^{-1} / (T_1T_{100})^{-1}\right)^2}{(T_1
T_{\rm As})^{-1}- (T_1T_{\rm ref})^{-1}}$ with $(T_1T_{\rm ref})^{-1}$, $(T_1T_{100})^{-1}$  as the
nuclear spin-lattice relaxation rates of ${\rm
LaO_{0.9}F_{0.1}FeAs}$ at arbitrary temperature and at $T_{\rm
ref}=100$ K, respectively, and 
with $(T_1 T_{\rm As})^{-1}$ as the  relaxation rate of ${\rm LaO_{0.9}F_{0.1}FeAs_{0.94}}$. Details of the fitting
procedure are described in the text.} 
\label{f.8}
\end{figure}

The dynamic spin susceptibility $\chi^{\prime\prime}({\bf
q},\omega)$ of ${\rm LaO_{0.9}F_{0.1}FeAs}$ and ${\rm
LaO_{0.9}F_{0.1}FeAs_{1-\delta}}$ samples was investigated by
$^{75}$As NMR spectroscopy. \cite{Hammer10,Grafe08} The nuclear
spin-lattice relaxation rate, $1/T_1T$, (where $T_1$ is the
nuclear spin-lattice relaxation time) is directly related to the
susceptibility $\chi^{\prime\prime}({\bf q},\omega)$ summed over
all $\bf q$ in the Brillouin zone.\cite{Penn96} Since the measurement of
$1/T_1T$ provides information about the electron-spin susceptibility
at all ${\bf q}$, we expect that 
for As-deficient
samples
the presence of  enhanced 
FM correlations characterized by ${\bf q}=0$ should also affect
their $1/T_1T$ rate.
The temperature
dependence of $1/T_1T$ for two typical ${\rm
LaO_{0.9}F_{0.1}FeAs}$ and ${\rm LaO_{0.9}F_{0.1}FeAs_{1-\delta}}$
samples is compared in Fig. \ref{f.8} (see also Ref.
\onlinecite{Hammer10}).
The inspection in Fig.\ \ref{f.8} shows that above $50K$
within the error bars the $1/T_1T$ of the reference samples
can be approximated by a linear  \emph{T}-dependence.  
Since in this temperature range the static susceptibility $\chi_s \propto T$
we can also expect that the $1/T_1T$ of the As-deficient sample can
be described by an equation similar to Eq. (\ref{3.4}),
if AVs do not effect essentially the AFM correlations. In this
case the \emph{T}-dependence of the ratio
$$
\frac{\left((T_1T_{\rm ref})^{-1} / (T_1T_{100})^{-1}\right)^2}{(T_1
T_{\rm As})^{-1}- (T_1T_{\rm ref})^{-1}}
$$
should be a straight line, with $(T_1T_{\rm ref})^{-1}$ and
$(T_1T_{100})^{-1}$ as the nuclear spin-lattice relaxation rates of
the reference samples at arbitrary temperature and at $T_{\rm
ref}=100$~K, respectively, and with $(T_1 T_{\rm As})^{-1}$  as the
relaxation rate of the As-deficient samples. From the
inset of Fig. \ref{f.8} it is seen that our experimental data can be well
described by this empirical fitting procedure. 
Therefore, we
conclude that the higher relaxation rates of the As-deficient
samples can be explained by a contribution related to the local moments
formed around AV. Since the contribution of the local moments on
the $1/T_1T$ measured on As nuclei is only part of the  total field
(itinerant electrons apart of AV interact with the As nuclei), the
estimated Curie temperature is somewhat lower, $\Theta_{\rm NMR}
\approx 7.5$ K as compared with the usual spin susceptibility
derived $\Theta \approx 18.8$ K obtained
for the static susceptibility (see Fig.\ \ref{f.6})
i.e.\ the hyperfine field for the As nuclei is less affected by the
local moments around the AV than the direct magnetic exchange interaction 
between the electrons.

\section{Superconducting properties: aspects of the Pauli limiting }

\begin{figure}[b!]
\includegraphics[width=9.5cm,angle=0]{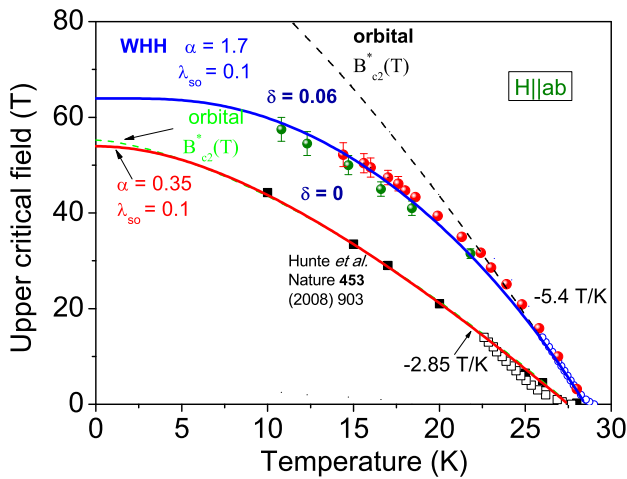}
\caption{(color online) Temperature dependence of the upper
critical field of ${\rm LaO_{0.9}F_{0.1}FeAs_{1-\delta}}$. Open
circles: DC data, closed circles: pulsed field measurements.
Green and red cloced circles stand for two different samples from the same batch
with the same As-deficiency
measured at the FZ Dresden and the IFW Dresden, respectively 
\cite{Fuchs10a, Fuchs10b}.
For comparison, the upper critical field data reported in
Ref.\ \onlinecite{Hunte08} are shown (closed squares). Open squares: DC data
of the reference sample measured at the IFW Dresden \cite{Fuchs10b}.
Solid lines: fit of the
experimental data to the WHH model. Dashed lines: the
orbital upper critical field $B^*_{c2}(T)$.
} 
\label{f.9}
\end{figure}

Up to now we discussed only high-temperature magnetic
properties of As-deficient samples in the 
metallic 
normal state phase. But also the 
superconductivity of these samples is  strongly
affected by the induced local magnetic moments. This concerns first of all the
\emph{T}-dependence of the upper critical field $B_{c2}$ of the
As-deficient samples shown in Fig. \ref{f.9}
(see also Refs.\ \onlinecite{Fuchs08,Fuchs09}).
(Here and below we ignore possible multiband effects\cite{Gurevich10}
for the sake of simplicity).   
For a polycrystalline sample
under consideration the $B_{c2}$ value refers to those grains
which are oriented with their \emph{ab} planes along the applied
field. Fig. \ref{f.9} demonstrates that the As-deficient samples
exhibit nearly two times higher slopes, $- dB_{c2}/dT$, at $T_c$
compared with the reference samples resulting in a very high
orbital upper critical field of $B_{c2}^*(0) = 106$ T.
(The high-field data of the As-stoichiometric optimally doped La-1111
were taken from Ref.\ \onlinecite{Hunte08}).
It was supposed previously
\cite{Fuchs08,Fuchs09} that AVs increase the disorder in FeAs
layers and reduce the mean free path \emph{l} of the conduction
electrons that results in a reduction of the effective coherence
length $\xi\sim (\xi_0l)^{1/2}$. However, the very narrow NQR
spectra of the As-deficient samples as compared with the 
As-stoichiometric
reference samples and the surprisingly stronger
$T$-dependence of the
nuclear spin-lattice relaxation rate in the superconducting state:
$T_1^{-1} \sim T^5$ (as compared
with $T_1^{-1} \sim T^3)$ for the reference sample) \cite{Hammer10,Grafe08}
indicate that actually our As-deficient samples might be even  
\emph{cleaner} than the reference samples, offering
this way a natural solution of 
the puzzling problem of
unusual or "supersmart impurities"
put forward in Ref.\ \onlinecite{Hammer10}.
In view of the repulsive interaction
between the As vacancies and an almost homogeneous distribution,
the role of remaining, possibly weak, disorder is rather unclear.
Therefore, the possible explanation of the reduction of the
BCS coherence length $\xi_0
\sim v^{}_{F}$ for the paired charge carriers in the As-deficient samples
might be alternatively explained at least partially by a decrease
of the Fermi velocities $v^{}_{F}$ of conduction electrons due to an
additional effective mass enhancement. This might be caused by enhanced 
magnetic correlations between itinerant electrons in the As-deficient
samples. In this respect further theoretical and experimental work
is required to elucidate the origin of such an enhancement.
Additionally, as it was pointed above, the correlated AVs probably
lead to a more uniform distribution of the F-dopants. This might
explain the slightly enhanced $T_c^{\rm As}=29$~K of the As-deficient
samples compared with $T_c=27.7$~K of the reference samples
\cite{Fuchs08,Fuchs09}, if the observed FM correlations 
in As-deficient samples
for some reason
do not impede the superconductivity.

On the other hand the resulting critical field at zero temperature
$B_{c2}^{\rm As}(0)$ of the As-deficient samples 
is strongly suppressed by paramagnetic spin effects 
\cite{Fuchs08,Fuchs09} and it may become comparable
with the
$B_{c2}^{\rm ref}(0)$ of the reference samples in spite of the mentioned
larger slope of $B_{c2}$ near $T_c$ and the improved $T_c$.
In general,
Pauli limiting behavior is closely related to an enhanced spin susceptibility
lowering the free energy in the normal state. In particular, the
condensation energy in the superconducting state at zero-field and
$T=0$ K is given by the free energy in the normal state at the
Pauli limiting field $B_p(0)$:
\begin{equation}\label{6.1}
\frac{1}{2}\chi_s^{\rm As}B_p^2(0) = \frac{B_c^2(0)}{8\pi},
\end{equation}
where $\chi_s^{\rm As}$ is defined in Eq. (\ref{3.6}).
Using Eq.\ (\ref{6.1}) and the ratio 
$\chi^{\rm As}_s/\chi^{\rm ref}_s > 6$
between the spin susceptibilities of the 
As-deficient and the reference samples the corresponding 
ratio $B_p^{\rm As}(0)$/$B_p^{\rm ref}(0)<0.4$ 
between the Pauli limiting fields of both samples
can be estimated. (Where we took 
into account that according to our
specific heat measurements, the $B_c(0)$-values 
of the As-deficient and reference samples are nearly the same.) On the 
other hand by fitting upper critical field data for As-deficient samples
to the curve predicted by the standard WHH model, the Pauli limiting field
$B_p^{\rm As}(0)$ = 114 T was estimated \cite{Fuchs08,Fuchs09} (see  Fig.\ 
\ref{f.9}) under the simplifying 
assumption that the spin-orbital scattering can be
neglected $\lambda_{s0} = 0$. 
In general for iron pnictides 
the effect of spin-orbit scattering on
$B_{c2}(T)$ is expected to be rather weak \cite{Fuchs09} 
and we adopt $\lambda_{s0} = 0.1$ as a more realistic value.
Then, the resulting
Pauli limiting field is limited by  $B_p^{\rm As}(0) = 88 T$. The same fitting
procedure yields
$B_p^{\rm ref}(0) > 200 T$ for the reference sample with  $\lambda_{s0} = 0.1$. 
In this case for the ratio between Pauli limiting fields of two
samples we have $B_p^{\rm As}(0)$/$B_p^{\rm ref}(0) <0.45$.
Therefore, using the 
experimentally measured susceptibilities the observed Pauli
limiting behavior can be explained at least qualitatively. 

\section{Defect and local moment aspects in Ba-122 systems grown from S\lowercase{n}-flux}

Finally, we would like to draw attention
that in particular the 
scenario of Pauli limited
superconductivity due to local moments 
proposed here might be applied also
to some K-doped Ba-122 pnictide superconductors.
For example, according to Refs.\ \onlinecite{Ni08,Mukhopadhyay08}, a
similar magnetic behavior as for our As-deficient samples reported here
has been observed including Pauli limiting \cite{Altarawneh08,Gasparov08}. 
However, at variance
with our findings also
a 20\% $T_c$-suppression and a
broadening of the NMR spectra has been observed
pointing to sizable disorder.
The peculiar 
magnetism 
has been 
ascribed by the authors
to large
local moments from a small amount of incorporated Sn occupying As-sites
\cite{Ni08,Mukhopadhyay08}.
This way being seemingly responsible for a significant paramagnetic 
pair-breaking and the observed $T_c$-suppression.
However, in our opinion a magnetic moment formation around Sn substitutions
for As sites seems to be
somewhat unlikely. Due to its strong interaction with the Fe-As host the 
formation of bound states as a prerequisite for magnetic moments is not 
expected and instead strong intra and interband scattering from a
non-magnetic impurity should occur. The latter can readily explain
the $T_c$ suppression within any multiband picture but especially for the 
$s_{\pm}$-pairing symmetry. Then the magnetic moments should be attributed
to As-vacancies as in the present work and/or to Fe residing outside
the Fe-As layer.
In these single 
crystals about {0.05} deficiency of As and about 0.03
excess of Fe have
been detected using
wavelength dispersive x-ray spectroscopy (WDX) \cite{Ni08}. 
Here, the formation of AV results probably
from the Sn-flux used in the crystal growth. The former acts as an 
As-getter analogously to the Ta-foil in our case and additionally partially
replace As atoms.
The resulting As deficiency is
comparable with that in our As-deficient
La-1111 samples.
In fact,     
the mentioned
phenomena can be explained semi-quantitatively
like in the present work.

\section{Conclusions}

We have measured and analyzed the static susceptibility 
together with the nuclear spin-lattice relaxation rate
$1/T_1T $ of As-deficient samples ${\rm
LaO_{0.9}F_{0.1}FeAs_{1-\delta}}$ $(\delta\approx 0.06)$ in
comparison with As-stoichiometric reference samples.
The concentration of As-vacancies Оґ has been first estimated from the 
EDX analysis and then a homogeneous
distribution and the amount of vacancies within 
the samples were confirmed 
and somewhat refined
by NQR measurements.
Quite remarkably the As-deficient samples show a
significant enhancement of the spin susceptibility $\chi_s^{\rm
As}/\chi_s^{\rm ref} \sim 3-7$. This enhancement provides
experimental evidence that the As vacancies in the La-1111
compound behave as magnetic defects with a net magnetic moment
associated within a $[{\rm V_{As}Fe_4}]$ or a $[{\rm V_{As}Fe_8}]$
complex defect about $m_{\rm eff} \approx 3.2\mu_{\rm B}$. The explanation of
this unusual effect is that the As vacancies induce a local spin
polarization of 3$d$-electrons near the Fermi energy. The enhanced
FM correlations between the conducting electrons are closely
related to a high value of the magnetic susceptibility in ${\rm
LaO_{0.9}F_{0.1}FeAs_{1-\delta}}$ via an enhanced Stoner factor. A
straightforward consequence of the enhanced spin susceptibility in
the As-deficient samples is that their upper critical field is
suppressed by spin-pair breaking at high external magnetic fields
achieved inevitably at low temperature. In contrast, the upper
critical field of the reference samples is not affected by
the spin-pair breaking and it can be described by the orbital
$B_{c2}(T)$ contribution down to the lowest temperatures. It would
be interesting to elucidate as well the microscopic origin of the
Pauli-limiting behavior reported also for other Fe-pnictide and
related selenide/telluride superconductors (see for example
\cite{Fuchs09,Fuchs10,Khim10}. This requires a detailed consideration of
the local electronic and magnetic structure of each defect type.
Possibly, the "local-moment" mechanism proposed here can be
applied with certain modifications also in those cases, especially
in the case of Se or Te-vacancies \cite{Khim10}. Investigations of
the spin susceptibility and NQR measurements of these compounds
are a necessary prerequisite, thus being of considerable interest.
In general, a better understanding of the new and complex
physics 
induced by various real defects present in many samples 
can also provide valuable insight into the superconducting mechanism itself, into the
role of correlation effects under debate, and into the complex interplay
with several competing magnetic and superconducting 
phases.

\begin{center}
{\bf ACKNOWLEDGEMENTS}
\end{center}
Discussions with K.\ Koepernik, R.\ Klingeler, D.\ Efremov, A,\ Yaresko, 
M.\ Kiselev, I.\ Eremin, M.\ Korshunov, 
M.\ Kulic, V.A.\ Gasparov,
J.\ M\'alek, S.\ Haindl, A. K\"ohler and
G. Behr$^\dag$ are grateful acknowledged. We thank P.\ Canfield for
critical discussion at an earlier stage of the present study. 
JvdB, SLD, and
KK (Grant No.\ BR4064/3-1), 
HJG (Grant No.\ GR3330/2), as well as SW and
BB (Grant No.\ BE1749/13) would like to thank the DFG Priority Programme
SPP1458 "High Temperature
Superconductivity in Iron Pnictides" 
for financial support. VG
would like to thank the Marie Curie Research Training Network
(RTN) NESPA, MRTN-CT-2006-035619, under the EU 6th Framework
Programme for financial support.


\begin{thebibliography}{99}
\bibitem{remark}After completing the present work we have learned, that similar enhanced
magnetism as reported here
caused by local magnetic moments of
"impurities" as well as Pauli limiting behavior have been observed for 
some
Ba-122 superconductors \cite{Altarawneh08,Gasparov08}. 
But the important interrelation  between the two phenomena and physical consequences have not  been mentioned.
(see also Sect.\  VII)
\end{thebibliography}
\end{document}